\documentstyle[12pt]{article}

\begin{document}
\title{\textbf{Anosov branches of dynamo spectra in one-dimensional turbulent plasmas}} \maketitle
{\textbf{L.C. Garcia de Andrade}-Departamento de F\'{\i}sica
Te\'orica-IF-UERJ-RJ, Brasil\\[-0.1mm]
\vspace{0.1cm} \paragraph*{Recently Guenther et al the globally diagonalized ${\alpha}^{2}$ dynamo operator spectrum [J Phys A 2007) in mean field media, and its Krein space related perturbation theory [J Phys A 2006). Earlier, an example of fast dynamos in stretch shear and fold Anosov maps have been given by Gilbert [PRSA [1993)). In this paper, analytical solutions representing general turbulent dynamo filaments are obtained in resistive plasmas. When turbulent diffusivity is present and kinetic helicity vanishes, a fast dynamo mode is obtained, and the Anosov eigenvalue obtained. The magnetic field lays down on a Frenet 2 plane along the filaments embedded in a 3D flow. Curvature effects on fast dynamo are also investigate. In case of weak curvature filaments the one dimensional manifolds in plasmas present a fast dynamo action. A parallel result has been obtained by Chicone et al [Comm Math Phys), in the case fast dynamo spectrum in two dimensional compact Riemannian manifolds of negative constant curvature, called Anosov spaces. While problems of embedding may appear in their case here no embedding problems appear since the one dimensional curved plasmas are embedded in three dimensional Euclidean spaces. In the examples considered here, equipartion between normal and binormal components of the magnetic field components is considered. In the opposite case, non Anosov oscillatory, purely imaginary, branches of the spectrum are found in dynamo manifold. Negative constant curvature non-compact $\textbf{H}^{2}$ manifold, has also been used in one-component electron 2D plasma by Fantoni and Tellez (Stat. Phys, (2008)). PACS: 47.65.Md. Key-word: dynamo plasma.}
\newpage
\section{Introduction}
 The simplest fast dynamo solution of a self induction equation of a stationary flow undergoing uniform stretching has been given by Arnold et al \cite{1}. Following this solution up to the present date many numerical simulations have been obtained \cite{2,3}. More recently Guenther et al \cite{4,5} investigated the spectra of the globally diagonalized ${\alpha}^{2}$ dynamo operator spectrum, by the examination of spherically symmetric eigenvalues. In this paper new analytical solutions of filamentary turbulent fast dynamos \cite{6} corresponding to a quasi Anosov magnetic dynamos. The word quasi refers to the fact that the corresponding stretching and squeezing eigenvalues are slightly different from the ones obtained by Arnold et al on the torus dynamo. In this paper the eigenvalue for a Anosov filament may be expressed in terms of curvature $\kappa$ by $\kappa[\frac{-1\pm\sqrt{5}}{2}]$ while the Anosov space eigenvalues are given by $\frac{3\pm\sqrt{5}}{2}$. Note that when the Frenet curvature of the one dimensional dynamo plasma is $\kappa=-1$, yields $\frac{1\mp\sqrt{5}}{2}$ which is a simply slightly change on stretching and squeezing properties of dynamos with respect to the Arnold et al eigenvalue. Another type of one dimensional plasma by Fantoni and Tellez \cite{7} in the realm of electron plasmas in 2D. Here, as happens in general relativity the plasma undergoes a Coriolis force which is given by the presence of the curvilinear coordinates effects present in the Riemann-Christoffel symbol in the MHD dynamo equation. In their non-relativistic plasma limit, this geometry has been used by Fantoni and Tellez in the context of plasma physics. They have used a Flamm's paraboloid, which is a non-compact manifold which represents the spatial Schwarzschild black hole, to investigate one-component two-dimensional plasmas. \newline
 Paper is organised as follows: Section II presents the mathematical formalism necessary to grasp the rest of the paper. In the next section the Frenet 2 plane magnetic fields across filamentary dynamo flows are presented. In section III the eigenvalues of the dynamo spectrum are obtained and the Anosov flow is obtained. Conclusions are presented in section IV.
\newpage
\section{One dimensional plasmas and 2 planes magnetic fields}
In this paper a mathematical technique based on one dimensional plasma in the form of filaments in the Euclidean three dimensional space $\textbf{R}^{3}$. Here instead of complications of Riemannian spaces the one dimensional plasmas leave on a flat space where only non vanishing curvature is the scalar one given by Frenet frame of the filaments. This frame is given by the three basis vectors $[\textbf{t},\textbf{n},\textbf{b}]$,
This frame vectors obey the following evolution equations
\begin{equation}
\frac{d\textbf{t}}{ds}={\kappa}(s)\textbf{n}
\label{1}
\end{equation}
\begin{equation}
\frac{d\textbf{n}}{ds}=-{\kappa}(s)\textbf{t}+{\tau}\textbf{b}
\label{2}
\end{equation}
\begin{equation}
\frac{d\textbf{b}}{ds}=-{\tau}(s)\textbf{n}
\label{3}
\end{equation}
Here ${\kappa}$ and ${\tau}$ are Frenet curvature and torsion scalars. The magnetic self-induction equation is written as
\begin{equation}
\frac{{\partial}\textbf{B}}{{\partial}t}={\nabla}{\times}({\alpha}\textbf{B})+
{\epsilon}{\Delta}\textbf{B}\label{4}
\end{equation}
where ${\alpha}=<\textbf{v}.{\nabla}{\times}\textbf{v}>$ represents the ${\alpha}$ helicity of the flow given by $\textbf{v}=v_{0}\textbf{t}$. Note that though one considers here that the modulus of the flow $v_{0}$ is considered as constant the flow is not necessarily laminar due to the dynamical unsteady nature of the frame vector $\textbf{t}$. This equation shall be expanded below, along the Frenet frame as
\begin{equation}
\textbf{B}(s,t)=B_{n}\textbf{n}+B_{b}\textbf{b}\label{5}
\end{equation}
Since in this section, one shall be considering the 2D case one shall assume from the beginning that the binormal component of the magnetic field $B_{b}$ shall vanish. Another simplification one shall addopt here is that the normal and binormal magnetic perturbations $B_{n}$ and $B_{b}$ shall be considered as constants. Some technical observations are in order now. The first is that this kind of Frenet frame used here, are called the isotropic Frenet frame, which considerers that even the frame is unsteady as here, in the sense that they depend upon time, the base vectors only depends upon the toroidal coordinate-s, and not other coordinates. Otherwise the Frenet frame is called anisotropic. The anisotropic may be more akin and suitable to turbulent phenomena but is much more involved, and shall be left to a next paper. Some simple use of the anisotropic Frenet frame can be found in simple Arnold like dynamos in reference \cite{10}. Let us start the MHD equations by the solenoidal divergence-free vector field by
\begin{equation}
{\nabla}.\textbf{B}=0 \label{6}
\end{equation}
as
\begin{equation}
{\partial}_{s}B_{b}-{\kappa}_{0}B_{n}=0 \label{7}
\end{equation}
Note that ${\kappa}_{0}={\tau}_{0}$ here represents the constant torsion ${\tau}_{0}$ equals the curvature ${\kappa}_{0}$ of the helical filamentary turbulence. By computing the relations
\begin{equation}
{\Delta}\textbf{t}=-{{\kappa}_{0}}^{2}\textbf{t}\label{8}
\end{equation}
\begin{equation}
{\Delta}\textbf{n}=-{{\kappa}_{0}}^{2}{\textbf{n}}\label{9}
\end{equation}
Since the 2D ${\alpha}^{2}$-dynamos is embedded in $\textbf{E}^{3}$, decomposition of the induction equation (\ref{4}), might be done along the three base vectors of the Frenet frame. By considering the Lyapunov chaotic behaviour of the magnetic field $|\textbf{B}|=B_{0}e^{{\lambda}t}$, the self induction equation is
\begin{equation}
{{\kappa}_{0}}B_{n}v_{s}=-{\kappa}_{0}B_{b}\label{10}
\end{equation}
\begin{equation}
{\gamma}B_{n}-{{\kappa}_{0}}B_{b}={\alpha}{\lambda}B_{n}-2{\beta}{{\kappa}_{0}}^{2}B_{n}\label{11}
\end{equation}
\begin{equation}
{\gamma}B_{b}-{{\kappa}_{0}}B_{n}v_{s}={\lambda}{\alpha}B_{n}-{\beta}{\kappa}_{0}B_{b}\label{12}
\end{equation}
Note that choice of 2-plane for the magnetic field polarised plane is transvected by the velocity of the Frenet curve.  This keeps some resemblance of the electric currents and the magnetic field which are orthogonal in general. This commonly happens in solid dynamos. The first dynamo expression, yields
\begin{equation}
B_{n}v_{s}=-B_{b}\label{13}
\end{equation}
which yields an expression for the rate at which normal component is transformed into the binormal one by the  velocity of one dimensional plasma flow. The remaining expressions can be collected in the form of a matrix to investigate the eigenvalue spectra
\begin{equation}
det[{\lambda}\textbf{I}-{D}_{\beta}]=0\label{14}
\end{equation}
Here $\textbf{I}$ represents the unit matrix
\vspace{1mm}
\begin{equation}$$\displaylines{\pmatrix{1&{0}\cr{0}&
1\cr}\cr}$$\label{15}
\end{equation}
where two-dimensional turbulent dynamo operator matrix $D_{\beta}$ can be written as
\begin{equation}
\vspace{1mm} $$\displaylines{\pmatrix{{\gamma}+2{\beta}{{\kappa}_{0}}^{2}-{\alpha}{\lambda}&{-{\kappa}_{0}}\cr
{-[{\alpha}{\lambda}+{{\kappa}_{0}}{v}_{s}]}&
{-[{\gamma}}+{\beta}{{\kappa}_{0}}^{4}]\cr}\cr}$$
\label{16}
\end{equation}
To simplify matters one shall choose the common practice in plasma physics of equipartition of magnetic field components in the form, $B_{n}=B_{b}$ which yields the following constraint on the flow, $v_{s}=-1$. Now within equipartition hypothesis, one shall address two cases of importance in dynamo physics: The first is the non turbulent ${\beta}=0$, laminar dynamo plasma case, which was also considered by Wang et al \cite{8} in the the cylindrical case without curvature or torsion. The second one when kinetic helicity vanishes while diffusive turbulence survives. In the first case the operator matrix ${D}$ reduces to
\begin{equation}
\vspace{1mm} $$\displaylines{\pmatrix{{\gamma}-{\alpha}{\lambda}&{-{\kappa}_{0}}\cr
{-[{\alpha}{\lambda}+{{\kappa}_{0}}{v}_{s}]}&
{-{\gamma}}\cr}\cr}$$\label{17}
\end{equation}
By taking the determinant of the matrix ${D}_{\alpha}$ one obtains the following algebraic second-order
\begin{equation}
{\gamma}^{2}+{\alpha}{\lambda}{{\gamma}}-[{\alpha}{\lambda}+{\kappa}_{0}]{\kappa}_{0}={0}\label{18}
\end{equation}
where one has considered that $v_{s}=-1$ and that the curvature is also very weak. Throghout the computations, the equipartition between curvature and torsion ${\kappa}_{0}={\tau}_{0}$ is assumed. By normalizing the curvature by ${\kappa}_{0}=1$ one obtains
\begin{equation}
{\gamma}_{\pm}= \frac{{\alpha}{\lambda}[-1\pm\sqrt{5}]}{2}\label{19}
\end{equation}
and since ${\alpha}{\lambda}=-1$ this expression finally reduces to
\begin{equation}
{\gamma}_{\mp}= \frac{[1\mp\sqrt{5}]}{2}\label{20}
\end{equation}
This eigenvalue characterizes Anosov spaces around torus manifold. Note that the first eigenvalue ${\gamma}_{-}\le{0}$, which indicates a non dynamo mode, while in the second case, ${\gamma}_{+}\ge{0}$, due to the expression
\begin{equation}
lim_{{\beta}\rightarrow{0}}\textbf{Re}{\gamma}(\beta)=0
\label{21}
\end{equation}
one obtains a fast dynamo mode. Here $\textbf{Re}$ represents the real part of the growth rate scalar ${\gamma}$. Since these eigenvalues represent the growth rate of the magnetic field, the plus sign represents the stretching of the filament while the minus sign represents the squeezing. Since stretching is fundamental for dynamo action , it is natural to find fast dynamos in the stretching mode. Note that if besides the non turbulent or laminar flow one also has a vanishing kinetic helicity only the marginal dynamo mode exists. In the second case, the turbulent dynamo operator ${D}_{\beta}$ becomes
\begin{equation}
\vspace{1mm} $$\displaylines{\pmatrix{{\gamma}+2{\beta}{{\kappa}_{0}}^{2}&{-{\kappa}_{0}}\cr
{-{{\kappa}_{0}}{v}_{s}}&
{-[{\gamma}}+{\beta}{{\kappa}_{0}}^{4}]\cr}\cr}$$
\label{22}
\end{equation}
Before computing the eigenvalues for this matrix it is easy to show that the turbulent dynamo filaments reduces to a laminar oscillating dynamo mode, since in the limit of vanishing ${\beta}$ this matrix now reduces to
\begin{equation}
\vspace{1mm} $$\displaylines{\pmatrix{{\gamma}&{-{\kappa}_{0}}\cr
{-{{\kappa}_{0}}{v}_{s}}&
{-{\gamma}}\cr}\cr}$$
\label{23}
\end{equation}
which upon computation of its determinant yields
\begin{equation}
{\gamma}_{\pm}= \pm{i{\kappa}_{0}}\label{24}
\end{equation}
Since and ${\kappa}_{0}\in \textbf{R}$, this expression represents an oscillatory non Anosov dynamo. In general the eigenvalue problem yields the following spectra
\begin{equation}
{\gamma}_{\mp}= \frac{[1\mp\sqrt{5}]}{2}\label{25}
\end{equation}
which is still Anosov in general.

\newpage
\section{Conclusions}
     Fast dynamos have been investigated in compact Riemannian manifolds, by Arnold et al \cite{1} and by Chicone and Latushkin \cite{4}. In this last case operator spectra in compact Riemannian spaces, have been determined. In this paper, fast kinematic are also investigated in the context of filaments in Anosov spaces \cite{10,11}. In this computation, Frenet holonomic frame is used, and the eigenvalue spectra is determined. Dynamical effects of the stretching of magnetic fields by plasma flow \cite{12} may appear elsewhere. All the physical applications make the model presented here, useful in physical realistic situations and deserve further study.
\section{Acknowledgements}
I am very much indebt to J-Luc Thiffeault, Dmitry Sokoloff, Yu Latushkin and Rafael Ruggiero for reading for helpful discussions on the subject of this work. I appreciate financial  supports from UERJ and CNPq.

  \end{document}